\documentclass[conference]{IEEEtran}
\IEEEoverridecommandlockouts
\usepackage{xcolor}
\usepackage{hyperref}
\hypersetup{
	colorlinks=true,
	linkcolor=black,
	citecolor=black,
	urlcolor=black
}
\usepackage{cite}
\usepackage{amsmath,amssymb,amsfonts}
\usepackage{algorithmic}
\usepackage{graphicx}
\usepackage{textcomp}
\usepackage{booktabs}
\usepackage{multirow}
\usepackage{tikz}
\usetikzlibrary{positioning,arrows.meta}
\def\BibTeX{{\rm B\kern-.05em{\sc i\kern-.025em b}\kern-.08em
    T\kern-.1667em\lower.7ex\hbox{E}\kern-.125emX}}
\begin{document}

\title{It's the Geometry, Not the Model: Effective Rank and Subspace Alignment in Functional Connectivity Classification
}
\author{Xiao Fan, Jingyuan Li, Yubo Han, Hongbin Guo, Guanya Li, Yang Hu, \\Wenchao Zhang, Weibin Ji, Yi Zhang\\ xiufan@stu.xidian.edu.cn }


\maketitle

\begin{abstract}
Resting-state functional connectivity (FC) is a widely used representation for classifying brain phenotypes and disorders, and most FC classification pipelines feed the full connectome into a model and place the burden of progress on the model. We instead argue that the geometry of FC itself is the dominant factor. Because FC is computed from whole-series correlations that are highly reliable within individuals and vary along relatively few directions across them, its across-subject variation is effectively low-dimensional, concentrating in a small leading effective subspace. Within a cohort, this implies that most of the nominal FC dimensions are redundant and that the discriminative structure available in the data is confined to a few directions. More importantly, across cohorts the orientation of this subspace need not agree, and such misalignment can drive cross-site transfer to fail even when the subspaces are of comparable size. Across 2,330 subjects from HCP, ABIDE, and ADHD-200, effective-rank analysis shows strong spectral concentration, and projecting onto only the leading effective-rank-scale components recovers most of the full-FC classification performance. On ABIDE, site-specific effective subspaces are only weakly aligned, and their principal-angle overlap predicts pairwise transfer after covariate adjustment, even though per-site effective ranks are comparable. Crucially, controlled rotations that perturb subspace orientation while preserving the mean and covariance spectrum drive transfer toward chance, whereas displacement-matched label-orthogonal rotations do not, isolating orientation as the factor that degrades transfer. Rather than proposing a new architecture, this study offers a diagnostic view of FC generalization, suggesting that cross-site evaluation and harmonization be judged by whether they align effective subspaces rather than only improve within-dataset accuracy.
\end{abstract}

\begin{IEEEkeywords}
functional connectivity, effective rank, representation geometry, subspace alignment, cross-site generalization
\end{IEEEkeywords}

\section{Introduction}
\label{sec:intro}
Functional connectivity (FC) derived from resting-state fMRI is widely used to characterize brain functional organization and to classify brain phenotypes and disorders.
Recent FC classification research has increasingly treated the high-dimensional connectome as a modeling problem, leading to graph convolutional networks (GCNs), connectome-specific convolutional neural networks (CNNs)~\cite{brainnetcnn}, and Transformer-based models~\cite{bnt}.
This line of work rests on a common assumption that FC contains discriminative structure which more expressive architectures can extract more fully, so progress is expected to come primarily from architecture design.

However, this model-centered framing does not explain two recurring observations in FC classification. First, simple multilayer perceptrons (MLPs), and even linear models, often match or exceed more elaborate graph-based architectures on standard FC classification tasks~\cite{han2026}. Second, FC classifiers remain fragile under cross-site transfer, a limitation that more sophisticated architectures do not resolve~\cite{klepl2026}. Together these observations raise a common representation-level question: how can the same FC geometry make decoding easy within a dataset yet fragile across sites?

Prior work has examined these observations only in isolation. Some studies show that functional-connectivity variation often has low-dimensional structure~\cite{kashyap2019,margulies2016}, and learning-theoretic results suggest that spectrally concentrated covariance can favor simple predictors~\cite{bartlett2020,tsigler2023}. Yet this literature treats strong within-dataset decoding mainly as a model-comparison issue and cross-site failure mainly as a domain-shift issue, leaving the two conceptually separate. We instead argue that both stem from a single representational property, the leading effective subspace of FC, that is, the small set of covariance directions that captures most across-subject FC variation.

This property also distinguishes FC from other neuroimaging inputs and clarifies why the representation, rather than the architecture, deserves attention. Standard medical images are raw high-dimensional signals for which expressive architectures are genuinely required to extract structure, and the architecture-centered framing above is largely inherited from that setting. FC, by contrast, is computed from whole-series correlations that are highly reliable within individuals and vary along relatively few directions across them~\cite{frame1}, so its across-subject variation concentrates in far fewer directions than its nominal edge count. 

Within a dataset, this leading effective subspace contains most of the discriminative FC signal, so most of the nominal FC dimensions are redundant and the discriminative structure is confined to a few directions. More importantly, across sites the same concentration becomes a transfer bottleneck. Effective rank sets the scale of the subspace, but transfer depends on whether site-specific effective subspaces are similarly oriented. When orientations diverge, a classifier trained on one site relies on directions that the other barely carries.

We evaluate this representation-geometry account on three public resting-state fMRI datasets, HCP, ABIDE autism, and ADHD-200, together spanning 2,330 subjects. Our claim is representation-centered. FC should not be treated merely as high-dimensional input, but as a low-effective-rank representation whose leading effective subspace shapes both within-dataset decoding and cross-site transfer. Our contributions are:
\begin{itemize}
	\item We reframe FC classification as an effective-subspace problem. 
	This view links two observations usually treated separately, strong within-dataset decoding by simple models and fragile cross-site transfer.
	
	\item We show that within-dataset FC decoding is largely supported by the leading effective subspace. 
	Effective-rank measurements and top-PC decoding localize most of the label-relevant FC signal to this subspace, where simple classical and shallow classifiers match or exceed more elaborate baselines.
	
	\item We show that the same effective subspace constrains cross-site transfer through its orientation. 
	Across ABIDE sites, principal-angle alignment predicts pairwise transfer after covariate adjustment, and controlled rotations of that subspace, at matched perturbation magnitude, isolate orientation as the factor that drives transfer down.
\end{itemize}

\section{Analysis Framework and Experimental Protocol}
\label{sec:method}

We organize the study around a representation-geometry question: why can FC be readily decoded within a dataset while remaining fragile under cross-site transfer? 
Our hypothesis is that both behaviors stem from the same effective-subspace geometry, which we make testable through the two quantities below.

\begin{figure}[t]
	\centerline{\includegraphics[width=0.92\linewidth]{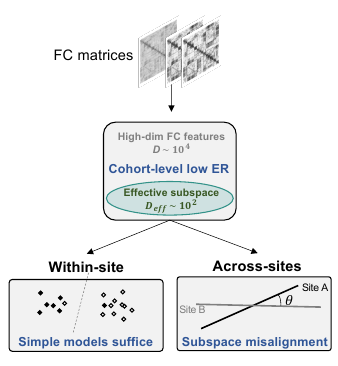}}
	\caption{The effective-rank and subspace-alignment analysis framework. 
		Vectorized FC features are nominally high-dimensional but may occupy a much smaller leading effective subspace. 
		The within-dataset branch tests whether this subspace contains most discriminative FC signal and explains the competitiveness of simple classifiers. 
		The cross-site branch tests whether site-specific effective subspaces are similarly oriented, using principal-angle overlap to quantify subspace alignment and relate it to transferability.}
	\label{fig:framework}
\end{figure}

We operationalize this account with two label-free geometric quantities (Fig.~\ref{fig:framework}). 
Effective rank measures the scale of the subject-level FC covariance, indicating how many covariance directions are effectively used. 
Principal-angle overlap measures the alignment between site-specific effective subspaces, indicating whether different sites represent FC variation along similar directions. 
Together, these quantities allow us to test two linked predictions of the same geometry: if discriminative FC signal is concentrated in the leading effective subspace, top principal component (PC) decoding should recover most full-FC performance and simple classifiers should be strong competitors; if cross-site transfer depends on subspace orientation, principal-angle overlap should predict transfer and controlled rotations of the leading subspace should degrade it. 
Robustness checks and raw blood-oxygen-level-dependent (BOLD) time series controls further rule out memorization, high dimensionality alone, relatedness, and label easiness as simpler explanations.

\subsection{Datasets and FC Feature Construction}
\label{sec:data}

We use three public resting-state fMRI datasets covering healthy and clinical phenotypes. 
For HCP sex classification, we retain 1,002 subjects from the S1200 release with available sex labels and use the Brainnetome atlas with $V=246$ regions~\cite{hcp,brainnetome}. 
In all tables and figures, HCP denotes this HCP sex-classification task. 
For ABIDE autism classification, we use 787 subjects from 22 sites. 
This full ABIDE cohort is used for within-dataset classification and effective-rank analyses. 
For cross-site subspace alignment, pairwise transfer, and leave-one-site-out (LOSO) analyses, we exclude the six sites with fewer than 20 subjects, leaving 16 sites with $n_s \geq 20$ for site-wise subspace estimation. 
The controlled orientation-rotation intervention (Section~\ref{sec:rotation-protocol}) additionally requires per-site split-half estimation and is therefore run on a further subset, the six largest sites ($n_s \geq 42$). 
These site sets are nested (22 sites for within-dataset analyses $\supseteq$ 16 sites for per-site cross-site analyses $\supseteq$ 6 sites for the rotation intervention), with each threshold set by the per-site sample the corresponding analysis needs for stable estimation.
For ADHD-200 classification, we use only the first session and retain subjects with valid AAL116 ROI time series, at least 80 time points, and a site label in the phenotypic table. 
We exclude sites with shorter time series, including Pittsburgh, WashU, and OHSU, leaving 541 subjects from 4 sites. 
ABIDE and ADHD-200 are both represented with the AAL atlas with $V=116$ regions~\cite{abide,adhd200,aal}. 

For each subject $i$, we compute the Pearson correlation matrix across ROI BOLD time series, apply Fisher's $z$-transform, and vectorize the upper-triangular entries excluding the diagonal to obtain $\mathbf{x}_i\in\mathbb{R}^D$, where $D=V(V-1)/2$. 
Stacking all subjects gives $X=[\mathbf{x}_1^\top,\ldots,\mathbf{x}_N^\top]^\top\in\mathbb{R}^{N\times D}$.
\subsection{Effective Rank of Subject-Level FC Covariance}
\label{sec:erank}

Effective rank quantifies the effective dimensionality of the subject-level FC covariance, which sets the scale of the leading effective subspace. 
Given the FC feature matrix $X$, we center it across subjects to obtain $\widetilde{X}$. 
Let $\lambda_i$ denote the nonzero eigenvalues of 
$\widetilde{X}^{\top}\widetilde{X}/(N-1)$. 
We compute the Shannon effective rank as
\begin{equation}
	\mathrm{ER}(X)=\exp\left(-\sum_i p_i\log p_i\right), 
	\qquad
	p_i=\frac{\lambda_i}{\sum_j \lambda_j}.
	\label{eq:erank}
\end{equation}
We adopt the Shannon (entropy-based) effective rank rather than a hard-threshold or numerical rank because it is a smooth, threshold-free functional of the eigenvalue spectrum that responds directly to how concentrated the spectrum is, which makes the resulting scale comparable across datasets and sites. A smaller $\mathrm{ER}$ indicates that across-subject FC variation is concentrated in fewer covariance directions, whereas a larger $\mathrm{ER}$ indicates that variation is spread across more directions.
\subsection{Within-Dataset Decodability and Representation Controls}
\label{sec:within-protocol}
We evaluate whether the leading effective subspace contains the discriminative FC signal that supports within-dataset decoding. 
We first perform top-$k$ principal-component decoding. 
Principal component analysis (PCA) is fitted on the training fold only, FC features are projected onto the leading $k$ components, and L2-regularized logistic regression (L2-LogReg) is evaluated on the held-out fold. 
We vary $k$ from small values to the $\mathrm{ER}$ and $2\mathrm{ER}$ scales, using rounded integer values for non-integer $\mathrm{ER}$. 
When reporting recovery relative to full FC, we use chance-normalized $\mathrm{AUROC}$ recovery,
\[
\frac{\mathrm{AUROC}_k-0.5}{\mathrm{AUROC}_{\mathrm{full}}-0.5}.
\]
We also include a mid-band control using components between $\mathrm{ER}$ and $2\mathrm{ER}$ to test whether performance is specific to the leading covariance directions rather than to dimensionality reduction alone. 
The upper bound is set to $2\mathrm{ER}$ so that this control spans a band of the same width as the leading effective subspace, giving a width-matched comparison between the leading $\mathrm{ER}$ directions and the next $\mathrm{ER}$ directions; it is chosen for this matched-control purpose rather than tuned.

We then compare classical baselines and representative neural architectures under shared preprocessing, splits, and metrics. 
The comparison covers classical linear and kernel models, shallow MLPs, graph-based neural networks, connectome-specific CNN-style architectures, and Transformer-based models. 
This design tests whether more elaborate architectures provide consistent gains over simple FC decoders when the leading effective subspace already contains most discriminative signal. 
Unless otherwise stated, evaluations use three random seeds and three stratified subject-level folds, and results are reported as mean$\pm$std over matched seed-fold runs. 
Hyperparameters for tunable models are selected by validation $\mathrm{AUROC}$ within each training fold.

We add three safeguards against simpler explanations. 
Label shuffling tests whether the pipeline can memorize arbitrary labels, random Gaussian features test whether high nominal dimensionality alone is sufficient, and family- versus subject-level CV on HCP tests whether relatedness drives performance.

Finally, we compare FC with raw BOLD time series from the same fMRI source. 
We evaluate flattened BOLD with ridge classification and structured BOLD with a compact temporal CNN under matched folds. 
This representation control asks whether simple FC classifiers succeed because the labels are intrinsically easy, or because FC makes much of the discriminative signal accessible to simple decoders.

\subsection{Cross-Site Subspace Alignment}
\label{sec:subspace-protocol}

Cross-site transfer requires more than within-site concentration of discriminative signal. 
If site-specific effective subspaces are poorly aligned, a classifier trained on one site may rely on directions that are weakly represented, or absent, in another. 
We test this idea on the 16-site ABIDE subset by estimating a site-specific effective subspace for each site.

For each site $s$, let $X_s$ denote its centered FC feature matrix. 
We obtain the top-$k$ principal components by singular value decomposition, with sensitivity evaluated over $k\in\{5,10,20\}$. 
For a pair of sites $(s,t)$, let $U_s,U_t\in\mathbb{R}^{D\times k}$ denote the corresponding orthonormal subspace bases. 
We define their principal-angle overlap as
\begin{equation}
	\mathrm{overlap}(s,t)
	=\frac{1}{k}\sum_{i=1}^{k}\cos^2(\theta_i)
	=\frac{1}{k}\left\|U_s^\top U_t\right\|_F^2,
	\label{eq:overlap}
\end{equation}
where $\theta_i$ are the principal angles obtained from the singular values of $U_s^\top U_t$. 
Larger overlap means the two sites represent FC variation along similar directions, whereas smaller overlap means these directions diverge, the geometric condition under which a source-site classifier leans on directions weakly represented at the target site and transfer is expected to degrade.

Pairwise transfer is measured by training L2-LogReg on site $s$ and testing on site $t$ for all ordered site pairs $(s,t)$, $s\neq t$. 
This gives a directional transfer matrix, whereas principal-angle overlap is symmetric. 
We therefore compute the Pearson correlation between overlap and transfer over ordered non-diagonal site pairs, and assess significance using a Mantel-style site-label permutation test with 5,000 permutations. 
In each permutation, site labels of the overlap matrix are jointly permuted along rows and columns, preserving the dyadic dependence structure while testing whether subspace alignment predicts directional transfer.
To ensure that the overlap--transfer association is not driven by site-level confounds, we residualize both overlap and transfer against five site-pair covariates using multiple regression quadratic assignment (MRQAP).
For each covariate we form a symmetric site-pair distance matrix: absolute differences in sample size, class balance (patient fraction), mean age, and sex ratio, together with the site-mean FC distance
\[
\left\|\bar{\mathbf{x}}_s-\bar{\mathbf{x}}_t\right\|_2,
\]
where $\bar{\mathbf{x}}_s$ is the mean FC vector of site $s$.
Overlap and transfer are regressed on these five covariate matrices, and the partial association between the overlap and transfer residuals is assessed by MRQAP with row--column permutations (5{,}000 permutations), which preserves the dyadic dependence structure.
Motion and mean framewise displacement were not included because they were unavailable in the ABIDE phenotypic table used here.

\subsection{Rank Control and Orientation Intervention}
\label{sec:rotation-protocol}

To separate subspace orientation from rank magnitude, we control for per-site effective rank in the dyadic analysis and repeat the comparison after matching site sample size. 
We then perform a controlled orientation-rotation intervention. 
For a target site $t$, let $\widetilde{X}_t$ denote its centered FC feature matrix and let $Q_\theta\in\mathbb{R}^{D\times D}$ be an orthogonal transformation that rotates selected FC directions by angle $\theta$. 
The rotated target representation is
\begin{equation}
	X_t^{(\theta)} = \mathbf{1}\bar{\mathbf{x}}_t^\top + \widetilde{X}_t Q_\theta,
	\qquad Q_\theta^\top Q_\theta=I .
	\label{eq:rotation}
\end{equation}
A classifier trained on the unrotated source site is then evaluated on $X_t^{(\theta)}$.

Because $Q_\theta$ is orthogonal and the original site mean $\bar{\mathbf{x}}_t$ is restored, the intervention preserves the feature mean and the covariance eigenvalue spectrum:
\begin{equation}
	\mathrm{cov}\!\left(X_t^{(\theta)}\right)
	= Q_\theta^\top \mathrm{cov}(X_t) Q_\theta .
	\label{eq:rotation_cov}
\end{equation}
Thus, the intervention perturbs orientation without changing lower-order structure. 
We rotate the leading effective subspace as the primary test and use displacement-matched discriminative and label-orthogonal rotations to distinguish label-relevant orientation effects from perturbation magnitude alone. Because this within-site design splits each site in half (one half trains the source classifier, the other is rotated and used as the target), it needs roughly twice the subjects of the whole-site analyses, so the rotation intervention is run on the six largest ABIDE sites ($n_s\geq42$), using 10 split-halves per site and seven rotation angles from $0^\circ$ to $90^\circ$. Effects are aggregated at the level of independent sites, giving a per-site paired comparison ($n=6$) between the discriminative and label-orthogonal arms.

\paragraph{Evaluation protocols.} Within-dataset analyses, including top-PC decoding, classifier comparisons, representation controls, and within-dataset robustness checks, use three stratified subject-level folds and three random seeds (seeds 1--3), yielding $3\times3$ seed--fold evaluations.  Results are reported as mean$\pm$std over these evaluations unless otherwise stated.  Cross-site analyses use site-level evaluation on the 16-site ABIDE subset.  Leave-one-site-out (LOSO) trains on all but one site and tests on the held-out site, whereas pairwise transfer trains on one site $s$ and tests on another site $t$ for all ordered pairs $(s,t)$, $s\neq t$.
\section{Results}
\label{sec:results}

\subsection{Low Effective Rank of FC Features}
\label{sec:erank-results}

We begin by establishing the structural property that underpins the within-dataset analysis. Effective-rank measurements across datasets are summarized in Table~\ref{tab:erank}. Although the FC feature vectors are nominally high-dimensional, their subject-level covariance is strongly concentrated: $\mathrm{ER}/D < 3\%$ for all three datasets. This means that across-subject variation in FC features is distributed over only a small number of effective covariance directions relative to the nominal feature dimension. The ratio $\mathrm{ER}/N$ ranges from $6\%$ to $32\%$, indicating that the effective dimensionality is also small relative to the available sample size. We therefore use $\mathrm{ER}/D$ and $\mathrm{ER}/N$ as empirical diagnostics of spectral concentration in each cohort.

\begin{table}[htbp]
\caption{Effective rank of vectorized FC features. $D$ is the nominal feature dimension, $\mathrm{ER}$ the Shannon effective rank of the subject-level FC covariance (Eq.~\ref{eq:erank}), and $\mathrm{ER}$/$D$, $\mathrm{ER}$/$N$ its ratios to the feature dimension and the sample size. $\mathrm{ER}$/$D<3\%$ on all three datasets indicates strong spectral concentration.}
\label{tab:erank}
\begin{center}
\begin{tabular}{lcccc}
\toprule
\textbf{Dataset} & \textbf{$D$} & \textbf{$\mathrm{ER}$} & \textbf{$\mathrm{ER}$/D (\%)} & \textbf{$\mathrm{ER}$/N (\%)} \\
\midrule
HCP & 30{,}135 & 62.8 & 0.21 & 6.3 \\
ABIDE & 6{,}670 & 108.5 & 1.63 & 13.8 \\
ADHD-200 & 6{,}670 & 174.2 & 2.61 & 32.2 \\
\bottomrule
\end{tabular}
\end{center}
\end{table}

\subsection{Top-PC Decoding in the Leading Effective Subspace}
\label{sec:topk-results}
\begin{figure}[t]
\centerline{\includegraphics[width=\linewidth]{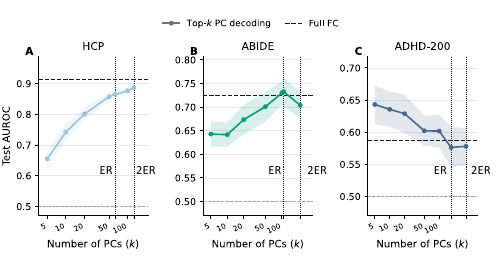}}
\caption{(A) HCP (B) ABIDE (C) ADHD-200. Top-$k$ PC classification of FC features. PCA is fitted within each training fold, and L2-LogReg is evaluated on held-out subjects after projecting FC features onto the leading $k$ principal components. Dashed horizontal lines show full-FC classification performance, and vertical dotted lines mark the rounded $\mathrm{ER}$ and $2\mathrm{ER}$ scales. The $\mathrm{ER}$-scale leading effective subspace recovers most full-FC performance. ADHD-200 reaches its peak before $\mathrm{ER}$, suggesting that its label-relevant signal is more concentrated than the covariance-defined $\mathrm{ER}$ scale.}
\label{fig:topk-pc}
\end{figure}

We next test whether the low-effective-rank structure of FC is directly relevant to supervised classification. After fitting PCA within each training fold, we project FC features onto the top-$k$ principal components and evaluate L2-LogReg on the held-out fold (Fig.~\ref{fig:topk-pc}). The leading effective subspace recovers most of the full-FC classification performance. At $k=\mathrm{ER}$, chance-normalized $\mathrm{AUROC}$ recovery is 88\% on HCP, 103\% on ABIDE, and 87\% on ADHD-200, so a subspace at the $\mathrm{ER}$ scale already reproduces most of the discriminative signal on all three datasets.

The three datasets differ in how the recovery curve behaves around this scale, and these differences follow from a single distinction. $\mathrm{ER}$ is computed from the unlabeled covariance spectrum and measures how widely across-subject variation is spread, which need not match how widely the label-relevant signal is spread. The two can therefore align, fall short, or slightly exceed one another. ABIDE is the aligned case: recovery reaches the full-FC level right at the $\mathrm{ER}$ scale (103\%), so the discriminative signal occupies about as many directions as the covariance does. ADHD-200 is the concentrated case: it has the largest $\mathrm{ER}$ (Table~\ref{tab:erank}) yet the weakest discriminative signal (Table~\ref{tab:arch}), because its across-subject variation is spread widely while the label-relevant part is confined to a few leading components. Recovery therefore peaks at a very small $k$ and then declines, since the remaining high-variance directions are largely label-irrelevant and adding them only injects noise. HCP is the mildly extended case: recovery is already 88\% at $\mathrm{ER}$ but rises to 94\% at $2\mathrm{ER}$, indicating that a small amount of label-relevant signal extends just beyond the $\mathrm{ER}$ scale while the bulk lies within it. This is consistent with our claim that the leading effective subspace carries most, not all, of the label-relevant signal.

The non-leading PC control confirms that this is a property of the leading directions rather than of dimensionality reduction alone. Components between $\mathrm{ER}$ and $2\mathrm{ER}$ yield $\mathrm{AUROC}$ values of 0.669, 0.545, and 0.508 on HCP, ABIDE, and ADHD-200, respectively, well below the leading-subspace performance. The label-relevant FC signal thus resides in the leading effective subspace, directly linking low-effective-rank geometry to the competitiveness of simple FC classifiers.

\subsection{Raw-BOLD Representation Control}
\label{sec:bold-results}
\begin{figure}[t]
\centerline{\includegraphics[width=\linewidth]{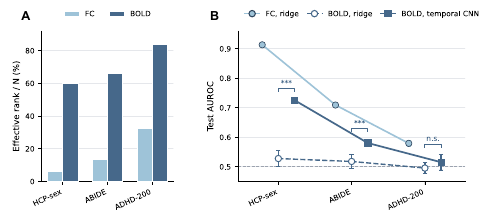}}
\caption{Representation choice shapes decoder complexity. 
	(A) Effective-rank-to-sample-size ratio for FC and raw BOLD representations. 
	(B) Test $\mathrm{AUROC}$ of FC with L2-LogReg, flattened raw BOLD with ridge classification, and raw BOLD with a temporal CNN. 
	Significance marks compare temporal CNN against flattened ridge over the $3\times3$ seed--fold runs.}
\label{fig:rank-interaction}
\end{figure}

Fig.~\ref{fig:rank-interaction} tests whether the competitiveness of simple FC classifiers is specific to the FC representation rather than to intrinsically easy labels. 
Raw BOLD and FC differ in a way that bears directly on their geometry. Raw BOLD is a time-resolved signal that fluctuates from frame to frame, so its discriminative structure is temporal and distributed rather than concentrated in a few static directions. FC, by contrast, summarizes each subject by whole-series correlations that are stable within individuals, which is why its across-subject variation collapses onto a small leading effective subspace. 
Consistent with this, using the same fMRI source, flattened raw BOLD with ridge classification remains weak, whereas a compact temporal CNN recovers signal on HCP and ABIDE. 
This contrast shows that discriminative information is present in the underlying fMRI data but is not equally accessible across representations, and that the low-dimensional structure is a property of how FC is constructed rather than of the parcellation or the labels. 
FC makes much of this information accessible to simple decoders, whereas raw BOLD requires temporal modeling, which is why analyzing the FC representation itself, rather than importing architectural complexity, is the appropriate lens here.

\subsection{Classifier Comparison under Low Effective Rank}

\label{sec:arch-results}

The effective-subspace result suggests that much of the FC decoding signal is already accessible to simple decoders. Table~\ref{tab:arch} evaluates this expectation using representative linear, kernel, MLP, connectome-specific, and Transformer-based models under shared splits and metrics.

Across HCP, ABIDE, and ADHD-200, simple FC decoders are consistently competitive. L2-LogReg achieves the best mean $\mathrm{AUROC}$ on HCP and ABIDE, while RBF SVM gives the best mean $\mathrm{AUROC}$ on ADHD-200. MLPs remain close to the best-performing models on HCP and ABIDE, but they are not uniquely superior. Representative graph, connectome-specific, and Transformer-based models do not provide consistent gains over these simpler baselines.

These results show that, for FC representations whose leading effective subspace already supports decoding, increasing architectural complexity is not automatically beneficial.

\begin{table}[htbp]
\caption{Architecture comparison. Test $\mathrm{AUROC}$ is reported as mean$\pm$std over $3\times3$ seed--fold runs. Boldface marks the best mean per dataset. L2-LogReg denotes L2-regularized logistic regression.}
\label{tab:arch}
\begin{center}
\scriptsize
\setlength{\tabcolsep}{2.2pt}
\renewcommand{\arraystretch}{1.05}
\begin{tabular}{llccc}
\toprule
\textbf{Family} & \textbf{Model} & \textbf{HCP} & \textbf{ABIDE} & \textbf{ADHD-200} \\
\midrule
Classical & L2-LogReg & \textbf{0.914$\pm$.002} & \textbf{0.724$\pm$.006} & 0.587$\pm$.009 \\
Classical & RBF SVM & 0.834$\pm$.009 & 0.670$\pm$.017 & \textbf{0.634$\pm$.006} \\
\midrule
Neural & MLP ($h=64$) & 0.896$\pm$.007 & 0.707$\pm$.011 & 0.594$\pm$.006 \\
Neural & MLP ($h=1024$) & 0.894$\pm$.005 & 0.710$\pm$.010 & 0.582$\pm$.014 \\
\midrule
Graph & GCN & 0.802$\pm$.008 & 0.632$\pm$.013 & 0.569$\pm$.016 \\
Graph & GAT-style & 0.807$\pm$.018 & 0.663$\pm$.006 & 0.574$\pm$.017 \\
Connectome & BrainNetCNN~\cite{brainnetcnn} & 0.846$\pm$.035 & 0.686$\pm$.008 & 0.596$\pm$.016 \\
Transformer & BNT~\cite{bnt} & 0.881$\pm$.006 & 0.676$\pm$.010 & 0.585$\pm$.018 \\
\bottomrule
\end{tabular}
\end{center}
\end{table}

\subsection{Robustness and Leakage Checks}
\label{sec:control-results}

The leading-effective-subspace analysis establishes the main link between low-effective-rank geometry and FC classification performance. 
The remaining within-dataset checks rule out simpler artifacts (Table~\ref{tab:controls}). 
Label shuffling reduces performance to chance, random Gaussian features with the same nominal dimensionality fail to reproduce true-label performance, and family-level cross-validation on HCP changes performance only marginally. 
Although the ADHD-200 Gaussian control is slightly above chance (0.545), it remains below the true-label reference (0.587). 
Together, these checks indicate that the observed FC performance is not explained by memorization, nominal dimensionality, or family relatedness.


\subsection{Cross-Site Subspace Alignment}
\label{sec:subspace-results}

\begin{figure*}[t]
\centerline{\includegraphics[width=0.92\textwidth]{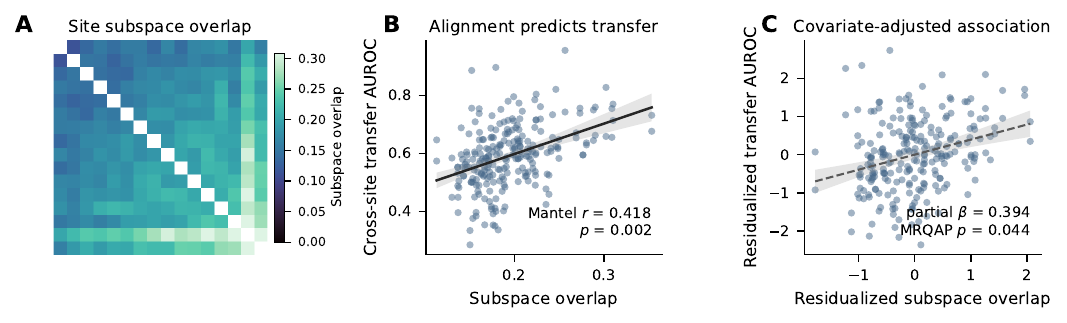}}
\caption{Cross-site subspace alignment on ABIDE ($k=10$). 
	(A) Principal-angle overlap between the 16 site-specific effective subspaces. 
	(B) Ordered pairwise transfer $\mathrm{AUROC}$ increases with subspace overlap; significance is assessed by Mantel-style site-label permutation. 
	(C) The overlap--transfer association remains after residualizing both axes against available site-pair covariates, including sample size, class balance, mean age, sex ratio, and site-mean FC distance.}
\label{fig:subspace}
\end{figure*}

We next examine whether cross-site transfer follows the alignment of site-specific effective subspaces. As a preliminary evaluation check, ABIDE leave-one-site-out (LOSO) cross-validation yields lower $\mathrm{AUROC}$ than random 3-fold cross-validation ($0.709\pm0.085$ vs. $0.724$), while remaining above chance. This shows that site structure affects evaluation, but that the classification signal is still detectable.
\begin{table}[]
	\caption{Control experiments and leakage checks. Reported values are mean test $\mathrm{AUROC}$. All controls are within-dataset.}
	\label{tab:controls}
	\begin{center}
		\small
		\setlength{\tabcolsep}{3pt}
		\begin{tabular}{lccc}
			\toprule
			\textbf{Control} & \textbf{HCP} & \textbf{ABIDE} & \textbf{ADHD-200} \\
			\midrule
			True labels (reference) & 0.914 & 0.724 & 0.587 \\
			Label shuffle & 0.514 & 0.491 & 0.495 \\
			Random Gaussian features & 0.531 & 0.535 & 0.545 \\
			\midrule
			Subject-level CV & 0.914 & -- & -- \\
			Family-level CV & 0.908 & -- & -- \\
			\bottomrule
		\end{tabular}
	\end{center}
\end{table}
We then quantify principal-angle overlap between the 16 ABIDE site-specific effective subspaces (Fig.~\ref{fig:subspace}A). At $k=10$, the mean overlap is 0.195, indicating substantial orientation differences across sites within the same ABIDE classification task. Higher overlap is associated with higher ordered pairwise transfer $\mathrm{AUROC}$ (Fig.~\ref{fig:subspace}B): Pearson $r=0.418$ with Mantel-style site-label permutation, $p=0.002$, with stable results across $k\in\{5,10,20\}$. 
The association remains after residualizing both overlap and transfer against site-pair differences in sample size, class balance, mean age, sex ratio, and site-mean FC distance (partial $\beta=0.394$, MRQAP $p=0.044$; Fig.~\ref{fig:subspace}C). Thus, site-specific effective subspaces that are more similarly oriented support better cross-site transfer.

We finally examine the role of effective-rank magnitude. Adding per-site effective rank to the dyadic regression preserves the overlap association ($\beta=0.500$, $p=0.032$). 
After subsampling every site to $n=20$, per-site effective ranks concentrate around $\approx 13$ with low cross-site variation (SD $\approx 1.2$). These controls show that, after sample-size matching, effective-rank magnitude is nearly uniform across sites, whereas subspace orientation remains informative for transfer.

\subsection{Controlled Subspace-Orientation Rotation}
\label{sec:rotation-results}

We next test whether subspace misalignment can directly degrade transfer, rather than merely correlate with it. We rotate selected FC directions in the target site while preserving the feature mean and covariance eigenvalue spectrum, and then evaluate a source-site classifier on the rotated target. This gives a controlled dose--response test of orientation mismatch.

\begin{figure}[t]
\centerline{\includegraphics[width=\columnwidth]{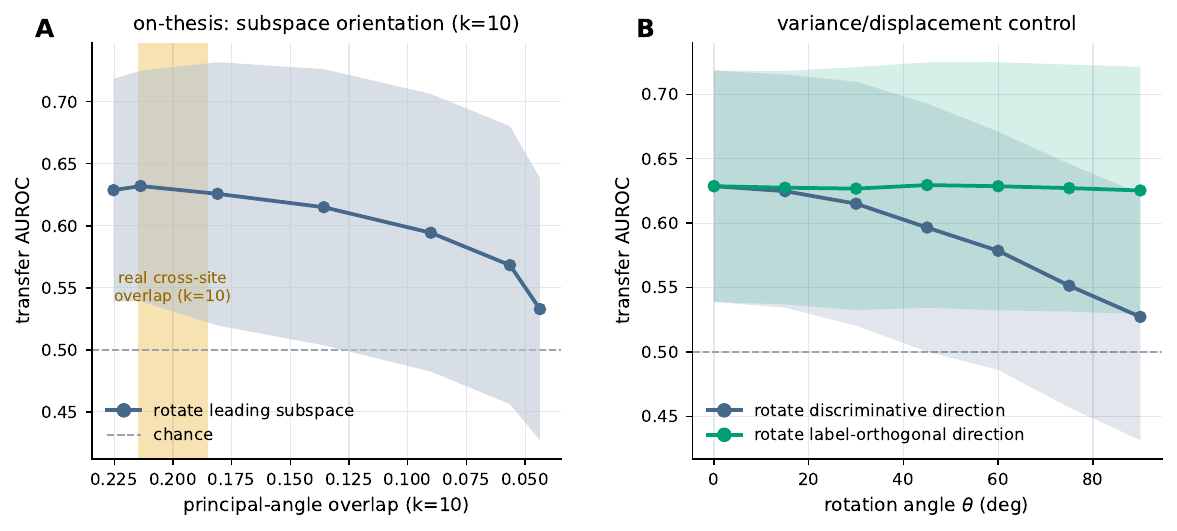}}
\caption{Controlled rotation of subspace orientation, on the six largest ABIDE sites ($n_s\geq42$; 10 split-halves per site, seven rotation angles $0^\circ$--$90^\circ$).
	(A) On-thesis test: rotating the leading effective subspace ($k=10$) drives source--target overlap down and degrades transfer $\mathrm{AUROC}$ monotonically toward chance; the shaded vertical band marks the empirical ABIDE overlap.
	(B) Variance/displacement control: rotating the discriminative direction degrades transfer, whereas rotating a label-orthogonal direction of matched displacement ($\lVert\Delta\mathbf{x}\rVert=6.82$) does not.
	The rotation is orthogonal and restores the site mean, preserving the feature mean and the covariance eigenvalue spectrum (changes $\sim10^{-15}$ and $\sim10^{-13}$), so only orientation is perturbed.
	Bands denote $\pm$SD across sites and split-halves; effect sizes and the per-site test are reported in the text.}
\label{fig:doseresponse}
\end{figure}

Rotating the leading effective subspace degrades transfer monotonically toward chance (Fig.~\ref{fig:doseresponse}A). As the rotation drives source--target overlap down across its realistic range, mean transfer $\mathrm{AUROC}$ falls from $\approx0.63$ at $\theta=0$ to $\approx0.53$ at $\theta=90^\circ$ (leading-subspace arm, $\Delta\approx0.096$); the swept band covers the empirical ABIDE overlap ($\approx0.195$ at $k=10$). Because the rotation is orthogonal and the site mean is restored, it leaves the feature mean (change $\sim7\times10^{-16}$) and the covariance eigenvalue spectrum (change $\sim2\times10^{-13}$) intact, so this decline reflects a change in orientation alone rather than any shift in mean or variance.

Matched-displacement controls show the effect is label-relevant rather than a generic consequence of perturbing the data (Fig.~\ref{fig:doseresponse}B). At equal displacement ($\lVert\Delta\mathbf{x}\rVert=6.82$), rotating the discriminative direction reduces transfer by $\Delta\approx0.10$ ($\mathrm{AUROC}\approx0.63\to0.53$), whereas rotating a label-orthogonal direction leaves it essentially flat ($\Delta\approx0.003$). Evaluated at the level of independent sites, the discriminative rotation lowers transfer at 5 of the 6 sites, against 0 of 6 for the label-orthogonal control (per-site paired $t$-test $p=0.034$; Wilcoxon $p=0.063$). With only six sites this is a controlled demonstration rather than a high-powered test; the population-level cross-site evidence remains the 16-site alignment--transfer association of Section~\ref{sec:subspace-results}. Taken together, the active--placebo dissociation under a mean- and spectrum-preserving rotation indicates that reorienting the leading effective subspace is, on its own, sufficient to degrade cross-site transfer, with perturbation magnitude held fixed.

\section{Discussion}
\label{sec:discussion}

\subsection{A Geometric Account of Two Behaviors}
The contribution of this study is not that FC has low effective rank, which is already known, but that a single geometric property, the leading effective subspace, accounts for two behaviors usually explained separately. Strong within-dataset decoding by simple models and fragile cross-site transfer are often treated as a model-comparison issue and a domain-shift issue, respectively. Our results tie both to the same subspace. Within a site it concentrates the discriminative signal, so simple decoders suffice and added architectural capacity has little structure left to exploit. Across sites the same concentration makes transfer depend on whether that subspace is similarly oriented, which the rotation intervention shows is sufficient, on its own, to drive transfer toward chance. Viewed this way, the two behaviors are not separate phenomena but two consequences of the same low-dimensional geometry.

\subsection{Implications for Future FC Studies}
This account is diagnostic rather than architectural, and it suggests concrete changes for how FC classification should be studied. First, model comparisons should be calibrated against the representation. When $\mathrm{ER}$-scale components already recover most full-FC performance, a strong simple baseline is the expected consequence of the representation, not a weak control, and an architecture should be credited only for signal it captures beyond the leading effective subspace. Otherwise, a comparison between a complex model and a simple classifier mainly reflects how much discriminative signal the FC representation has already exposed.

Second, cross-site evaluation should be treated as a geometric question. Poor transfer should not be attributed by default to model capacity or generic domain shift; when site-specific subspaces are misaligned, the source-site decision directions are simply under-expressed at the target. We therefore suggest that harmonization and domain-adaptation methods be assessed by whether they increase subspace alignment, not only by whether they raise within-dataset $\mathrm{AUROC}$, and that principal-angle overlap serve as a lightweight diagnostic for anticipating transfer before models are trained. Traveling-subject or harmonized multi-site cohorts would allow this alignment view to be tested where acquisition differences are controlled.

These implications are representation-specific. The raw-BOLD control shows that when a representation does not expose the discriminative signal to a linear model, temporal modeling still recovers information that linear decoding misses, so complex models remain necessary in general. For vectorized FC, however, the leading effective subspace is the object that governs both model comparison and transfer, and treating it as such is more informative than adding architectural complexity alone.
\subsection{Limitations}
Two limitations bound these conclusions. First, the cross-site analysis adjusts for sample size, class balance, mean age, sex ratio, and site-mean FC distance, but not for motion or mean framewise displacement, which were unavailable in the phenotypic table used here; with 16 sites the adjusted dyadic analysis also has limited power. Second, the rotation intervention is a controlled synthetic perturbation on the six largest sites, so its strength lies in the mean- and spectrum-preserving active-placebo dissociation rather than in the per-site sample size. Natural site differences are not purely rotations, and traveling-subject or harmonized multi-site cohorts would be needed to test whether the same subspace mechanism holds once population and acquisition differences are tightly controlled.

\section{Conclusion}
\label{sec:conclusion}

FC classification is strongly shaped by representation geometry. Within a dataset, the leading effective subspace contains much of the discriminative FC signal, allowing simple linear and shallow classifiers to compete with graph, connectome-specific, and Transformer architectures. 
Across sites, the same concentration becomes a vulnerability: transfer depends on whether different sites orient that subspace the same way, and controlled orientation perturbations degrade transfer toward chance. Together these results trace both behaviors, strong within-dataset performance of simple models and fragile cross-site transfer, to a single geometric origin. FC should therefore be interpreted not merely as a high-dimensional connectome input, but as a low-effective-rank representation whose leading subspace governs decoding and generalization alike.

%
\section*{Code Availability}
Anonymized code and figure scripts are available at:
\href{https://anonymous.4open.science/r/LowRank_A-CBA0/}
{https://anonymous.4open.science/r/LowRank\_A-CBA0/}.

\bibliographystyle{IEEEtran}
\bibliography{references}
\end{document}